# RSS-based LTE Base Station Localization Using Single Receiver in Environment with Unknown Path-Loss Exponent

Suhui Jeong
School of Integrated Technology
Yonsei University
Incheon, Korea
ssuhui@yonsei.ac.kr

Halim Lee
School of Integrated Technology
Yonsei University
Incheon, Korea
halim.lee@yonsei.ac.kr

Taewon Kang
School of Integrated Technology
Yonsei University
Incheon, Korea
taewon.kang@yonsei.ac.kr

Jiwon Seo
School of Integrated Technology
Yonsei University
Incheon, Korea
jiwon.seo@yonsei.ac.kr

*Abstract*—With the increasing demand for location-based services, localization technology research has recently intensified. Received signal strength (RSS)-based localization has the advantage of simplicity. However, as RSS-based localization requires the path-loss model parameters, it is difficult to use in place on which those parameters are unknown. In prior research, a transmitter localization algorithm with multiple stationary receivers was proposed for use under unknown path-loss exponent (PLE) conditions. However, if a mobile receiver is utilized, the localization would be possible with a single receiver alone. In this paper, we suggest a method of RSS-based LTE base station (BS) localization with a single mobile receiver when the PLE is unknown. We also propose an efficient mobile-receiver movement method to improve the PLE estimation and BS localization accuracy. Simulation results demonstrate the performance of the proposed methods.

## I. Introduction

In the recent years, mobile positioning has become important with the expansion of location-based services. The global navigation satellite system (GNSS) technology, including the Global Positioning System (GPS) of the United States, is the most popular positioning technology, owing to its high positioning accuracy [1-3]. However, because of the weak intensity of the signal, it is vulnerable to not only man-made attacks such as signal jamming but also ionospheric anomalies [4-8]. Similarly, in urban areas, GNSS accuracy is often reduced significantly by multipaths, where signals are reflected by buildings, and non-line-of-sight (NLOS) conditions, where signals are obscured by buildings. To protect against such signal reception degradation and low-performance situations, many studies on alternative navigation are underway, such as long-range navigation (Loran) [9-11] and distance measuring equipment (DME) [12].

Signals such as long term evolution (LTE) signals are good alternatives to complement or replace these shortcomings of GNSS. The LTE signals have the advantage that they can be received at high intensities with high information transmission rates and have wide radio bandwidths. In addition, the LTE infrastructure has already been established; therefore, there are no additional infrastructure installation costs. However, it is necessary to obtain the location information of nearby LTE base stations (BSs) to enable LTE-based positioning. Since the BS location information in the existing database has low accuracy, BS location survey should be performed by a BS localization technique. Direct access of a BS is not always possible. For example, it may not be possible to access a BS on a rooftop of a private building. Thus, a remote localization method is desired.

Several types of localization techniques have been proposed [13-15]. For instance, there are ways to use the time of arrival (TOA) [16], angle of arrival (AOA) [17], and received signal strength (RSS) [18-22]. The RSS-based techniques do not require time synchronization; therefore, they are more straightforward for use [23]. Thus, we consider the RSS-based localization in this paper. However, the RSS-based localization requires the information about the path-loss exponents (PLEs).

A technique for BS localization using signals of opportunity (SOPs) and mobile receivers has been developed [24]. This method has high localization accuracy, but it requires multiple mobile receivers to localize the BS.

In this paper, we propose a method based on [23], for estimating the BS location using RSS with only one mobile receiver in conditions where the PLE is unknown. This method is more convenient and economic than the method using multiple mobile receivers. The way the mobile receiver moves will affect the performance of the proposed method. Therefore, we also propose a method to move the mobile receiver in a suitable manner to improve the BS localization performance.

## II. PLE Estimation Method

The BS coordinate that we want to estimate is $(x, y)$ and the mobile receiver's coordinate that we already know is $(p, q)$. The general model for the measured RSS and distance between the BS and a mobile receiver is expressed as [23]:

$$R = R_0 - 10n \log\left(\frac{d}{d_0}\right) + X \quad (1)$$

where $R$ [dBm] is the measured RSS of the receiver at a distance $d$ [m] from the BS, $R_0$ [dBm] is the RSS at the reference distance $d_0 = 1$ m, $n$ is the PLE, and $X$ is the shadow noise, which is a zero-mean Gaussian random variable with standard deviation σ [dB]. $R$ can also be modeled as a Gaussian random variable, as $R_0$ and $d_0$ are constants. Thus, the probability density function of $R$ according to the distance $d$ is represented by [23]:

$$f\left(\frac{R}{d}\right) = \frac{1}{\sigma}\exp\left[-\frac{\left(R-\left(R_0-10n\log\left(\frac{d}{d_0}\right)\right)\right)^2}{2\sigma^2}\right]. \quad (2)$$

Therefore, the maximum-likelihood estimator of $d$ is [23]:

$$\hat{d} = d_0 10^{\frac{(R_0-R)}{10n}}. \quad (3)$$

We can modify this expression as [23]:

$$\hat{d} = d_0 \left(\frac{10^{\frac{R_0}{10}}}{10^{\frac{R}{10}}}\right)^{\frac{1}{n}}. \quad (4)$$

It is assumed that three or more receivers at known locations are available in [23] to localize the transmitter, but we use only one mobile receiver in this paper. Thus, the receiver index used in [23] is dropped in (1) through (4).

As in [23], we substitute $10^{\frac{R_0}{10}} = P_0$ and $10^{\frac{R}{10}} = P$ to simplify (4).

$$\hat{d} = d_0 \left(\frac{P_0}{P}\right)^{\frac{1}{n}}. \quad (5)$$

Unlike [23], we assume that $R_0$ is known from the database and thus $P_0$ is also known. Using $\hat{d} \approx \sqrt{(p-x)^2 + (q-y)^2}$ and $d_0 = 1$, (5) can be rearranged as:

$$2P^{\frac{2}{n}}px + 2P^{\frac{2}{n}}qy - P^{\frac{2}{n}}(x^2 + y^2) = P^{\frac{2}{n}}(p^2 + q^2) - P_0^{\frac{2}{n}}. \quad (6)$$

As in [23], we substitute $x^2 + y^2 = S$ and represent (6) in a matrix form.

$$\mathbf{A}\boldsymbol{\theta} = \mathbf{b}, \quad (7)$$

where

$$\mathbf{A} = \begin{bmatrix} 2P^{\frac{2}{n}}p & 2P^{\frac{2}{n}}q & -P^{\frac{2}{n}} \end{bmatrix}, \mathbf{b} = \begin{bmatrix} P^{\frac{2}{n}}(p^2+q^2) - P_0^{\frac{2}{n}} \end{bmatrix},$$
$$\boldsymbol{\theta} = \begin{bmatrix} x \\ y \\ S \end{bmatrix}. \quad (8)$$

Multiple stationary receivers are assumed in [23], but this paper assumes one mobile receiver. When the receiver is in motion, new measurements at new locations become available. Those new measurements can be represented as additional rows of the **A** and **b** matrices as in (9).

$$\mathbf{A} = \begin{bmatrix} 2P_1^{\frac{2}{n}}p_1 & 2P_1^{\frac{2}{n}}q_1 & -P_1^{\frac{2}{n}} \\ 2P_2^{\frac{2}{n}}p_2 & 2P_2^{\frac{2}{n}}q_2 & -P_2^{\frac{2}{n}} \\ \dots & \dots & \dots \\ 2P_i^{\frac{2}{n}}p_i & 2P_i^{\frac{2}{n}}q_i & -P_i^{\frac{2}{n}} \end{bmatrix},$$
$$\mathbf{b} = \begin{bmatrix} P_1^{\frac{2}{n}}(p_1^2+q_1^2) - P_0^{\frac{2}{n}} \\ P_2^{\frac{2}{n}}(p_2^2+q_2^2) - P_0^{\frac{2}{n}} \\ \dots \\ P_i^{\frac{2}{n}}(p_i^2+q_i^2) - P_0^{\frac{2}{n}} \end{bmatrix} \quad (9)$$

where $(p_i, q_i)$ is the $i$th position of the mobile receiver. Since $P_0$, $P_i$, and $(p_i, q_i)$ are known, the PLE, $n$, is the only unknown parameter to be solved to obtain the BS coordinate $(x, y)$.

Having different PLE values depending on the environment, makes it difficult to estimate $n$ for an anonymous environment. Fortunately, there is always an upper and lower bound for $n$ [25]. This helps us to search for the optimal value of PLE, $n_{opt}$, within the boundary.

There are two ways to obtain the estimated distance, $\hat{d}$, between the BS at $(x, y)$ and the mobile receiver at $(p, q)$. The first way is to use (5) with the measured RSS value of the mobile receiver and an assumed value of PLE, which is $n_j$ within the range of $N_{min} \le n_j \le N_{max}$. The second way is to use (7) and (9). We can substitute $n_j$ in (9) and then use the linear least-squares estimator to obtain $\widehat{\boldsymbol{\theta}}_{n_j} = \left[\hat{x}_{n_j}, \hat{y}_{n_j}, \hat{S}_{n_j}\right]^T$. The estimated distance, $\hat{d}$, is the distance between the coordinates of the BS and the mobile receiver, which are $(\hat{x}_{n_j}, \hat{y}_{n_j})$ and $(p, q)$, respectively.

Note that those two distance estimates for a given $n_j$ from the two different estimation methods should be identical if $n_j$ represents the true value of PLE. As in [23], we set $N_{min} = 1$ and $N_{max} = 5$. Then, $n_j$ values within the range were tried with a step size of 0.1. The $n_{opt}$ value was selected as the $n_j$ value that minimizes the difference between the two distance estimates explained above. If we decrease the step size, the estimation accuracy of $n_{opt}$ increases but the computational time also increases.

## III. SIMULATION RESULT

Because a mobile receiver collects information as it moves, the way it moves affects the localization performance. We simulated two exploration methods of the mobile receiver: in the first method, the mobile receiver moved randomly, and in the second, our proposed method, the mobile receiver moved to the nearest corner and visited every corner around the edge of the map to increase geometric diversity.

Simulations were conducted to verify the performance of the proposed exploration algorithm and the PLE estimation method. The simulation environment was 45 m × 45 m, with a 1 m grid. In each trial, the BS and mobile receiver were positioned randomly and operated for 150 s. The trial was conducted 100 times with different values of PLE, $R_0$, and σ.

We evaluated the effects of the path-loss model parameters (i.e., true PLE, $R_0$, and σ) on the PLE estimation accuracy and the BS localization error as follows.

## A. Comparison According to PLE

The PLE estimation error according to the random receiver movement remains larger than the case of the proposed strategic movement (Table I). The BS localization error also remains higher than the case of the proposed method (Table II and Fig. 1).

TABLE I. AVERAGE ESTIMATED PLE ($n_{opt}$) FOR EACH TRUE PLE VALUE ($R_0 = -27 dBm, \sigma = 3dB$)

| True PLE, n | 2 | 2.5 | 3 | 3.5 | 4 |
|---|---|---|---|---|---|
| Random | 4.0 | 4.5 | 4.8 | 4.8 | 4.2 |
| Proposed method | 2.2 | 2.7 | 3.2 | 3.7 | 4.1 |

TABLE II. AVERAGE ROOT MEAN SQUARE (RMS) BS LOCALIZATION ERROR (m) FOR EACH TRUE PLE VALUE ($R_0 = -27\ dBm, \sigma = 3dB$)

| True PLE, n | 2 | 2.5 | 3 | 3.5 | 4 |
|---|---|---|---|---|---|
| Random | 24.4 | 23.7 | 23.5 | 20.8 | 11 |
| Proposed method | 10.6 | 9.3 | 7 | 6.5 | 5.0 |

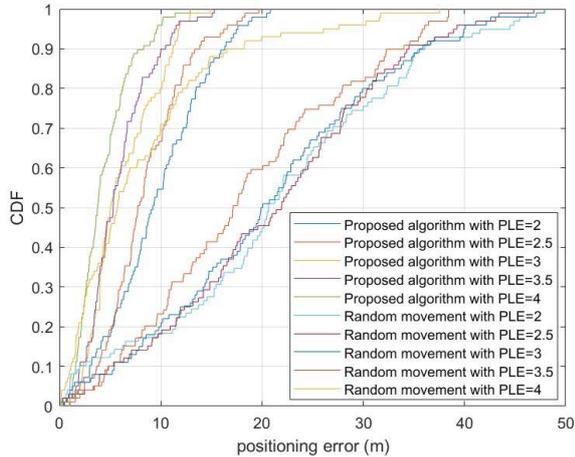

Fig. 1. Cumulative distribution function of BS localization error.

## B. Comparison According to Shadow Noise

As the shadow noise increases, the PLE estimation and BS localization errors of both the random movement and proposed method increase. Whereas, the proposed method demonstrated less errors than the random movement case as presented in Tables III and IV and Fig. 2.

TABLE III. AVERAGE ESTIMATED PLE ($n_{opt}$) FOR EACH SHADOW NOISE VALUE (TRUE PLE = 3, $R_0 = -27\ dBm$)

| Shadow noise, σ [dB] | 1 | 2 | 3 | 4 |
|---|---|---|---|---|
| Random | 3.4 | 4.6 | 4.8 | 4.8 |
| Proposed method | 3 | 3.1 | 3.2 | 3.3 |

TABLE IV. AVERAGE RMS BS LOCALIZATION ERROR (m) FOR EACH SHADOW NOISE VALUE (TRUE PLE = 3, $R_0 = -27\ dBm$)

| Shadow noise, σ [dB] | 1 | 2 | 3 | 4 |
|---|---|---|---|---|
| Random | 10.5 | 23.2 | 23.5 | 22.2 |
| Proposed method | 1.0 | 4.5 | 7 | 10.2 |

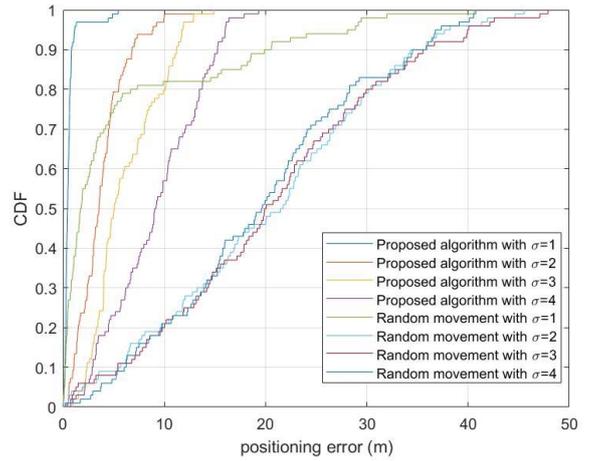

Fig. 2. Cumulative distribution function of BS localization error.

## C. Comparison According to $R_0$

When the $R_0$ increases, the PLE estimation errors in both the random and proposed methods show small variations; but, the random method error is greater (Table V). Also, the BS localization error of the proposed method is significantly smaller than that of the random movement (Table VI, Fig. 3).

TABLE V. AVERAGE ESTIMATED PLE ($n_{opt}$) FOR EACH $R_0$ VALUE (TRUE PLE = 3, $\sigma = 3dB$)

| $R_0$ [dBm] | −20 | −25 | −30 | −35 | −40 |
|---|---|---|---|---|---|
| Random | 4.4 | 4.7 | 4.8 | 4.8 | 4.4 |
| Proposed method | 2.7 | 3.0 | 3.4 | 3.7 | 4.0 |

TABLE VI. AVERAGE RMS BS LOCALIZATION (m) FOR EACH $R_0$ VALUE (TRUE PLE = 3, $\sigma = 3dB$)

| $R_0$ [dBm] | −20 | −25 | −30 | −35 | −40 |
|---|---|---|---|---|---|
| Random | 23.6 | 23.5 | 24.8 | 21.4 | 17 |
| Proposed method | 7.6 | 7.5 | 7.9 | 7.1 | 7.6 |

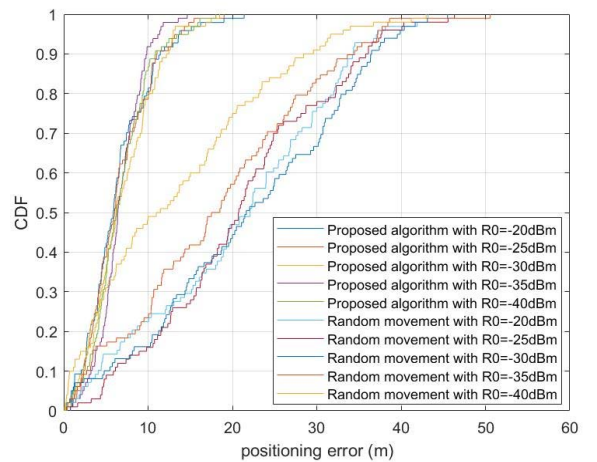

Fig. 3. Cumulative distribution function of BS localization error.

## IV. CONCLUSION

In this paper, an RSS-based BS localization method was proposed with a mobile receiver for cases where the PLE was unknown. We searched for an optimal PLE with a single mobile receiver. Then, the location of the BS was estimated. We also suggested the strategic movement of the mobile receiver to improve the PLE estimation and BS localization accuracy. The performance of the proposed methods were demonstrated through simulations.

ACKNOWLEDGMENT

This work was supported by Institute for Information & Communications Technology Planning & Evaluation (IITP) grant funded by the Korea government (KNPA) (No. 2019-0-01291, LTE-based accurate positioning technique for emergency rescue). This research was also supported by the HPC Support Project funded by the Ministry of Science and ICT (MSIT) and the National IT Industry Promotion Agency (NIPA) of Korea.